\documentclass[conference]{IEEEtran}

\usepackage{array}
\usepackage{balance}
\usepackage{cite}
\usepackage{graphicx}
\usepackage{url}

\title{LLM-Based Static Verification of Code Against Natural-Language Requirements: An Industrial Experience Report}

\author{
\IEEEauthorblockN{Zhi Quan Zhou}
\IEEEauthorblockA{
\textit{NIO Inc.}\\
China\\
george.zhou@nio.com}
\and
\IEEEauthorblockN{Dave Towey}
\IEEEauthorblockA{
\textit{University of Nottingham Ningbo China}\\
China\\
Dave.Towey@nottingham.edu.cn}
\and
\IEEEauthorblockN{Tsong Yueh Chen}
\IEEEauthorblockA{
\textit{Swinburne University of Technology}\\
Australia\\
tychen@swin.edu.au}
}

\begin{document}

\maketitle

\begin{abstract}
Large language models (LLMs) are increasingly used to generate requirements specifications, design documents, code, and test cases. In contrast, much less attention has been given to a more difficult assurance problem: statically verifying whether implemented code satisfies requirements written in natural language. Conventional static analysis tools such as Coverity and SonarQube are effective at detecting coding defects and known vulnerability patterns, but they cannot determine whether program behavior matches intended business logic. For example, a program intended to perform multiplication but implemented using addition may contain no memory-safety flaw or recognizable vulnerability signature. Detecting such defects requires reasoning over the specification rather than the code alone. Software testing can expose some of these mismatches, but its effectiveness depends heavily on test design, executable artifacts, and runtime environments. Many business-logic defects remain difficult to detect unless corner cases or large input spaces are extensively exercised.

This article presents a two-stage LLM-based workflow for addressing this challenge in an intelligent-vehicle cybersecurity case study. In the first stage, an AI-based rule miner extracts verifiable rules from natural-language requirements while explicitly identifying ambiguity, self-contradiction, and other non-verifiable statements. The completeness of the extracted rules are enhanced by using a simple metamorphic relation. In the second stage, an AI-based code auditor checks implementation evidence against the extracted rules. This separation is intentional. Instead of asking a single LLM to directly verify code against lengthy natural-language specifications, the workflow introduces a structured intermediate representation to reduce hallucination, output variability, limited explainability, and context loss. The resulting approach can be viewed as a requirement-aware and semantics-aware form of static analysis that complements rather than replaces software testing. By analyzing requirements and source code without requiring compilation, execution, or runtime environments, the method shifts a substantial portion of verification and validation activities left in the development lifecycle. Public-safe case materials suggest that combining requirement clarification with business-logic-oriented code verification is practically useful and can verify more than 50\% of the requirements that previously could only be evaluated through software testing. This research also highlights that LLM-based static analysis is a new approach to addressing the test oracle problem.
\end{abstract}


\section{Introduction}

LLMs are now routinely used across the software lifecycle, including requirements drafting, design support, code generation, and test generation. More broadly, the use of natural language processing for requirements engineering is well established, although ambiguity handling, formalization, and traceability remain persistent challenges~\cite{dalpiaz2018,zhao2021}. Much less work, however, has examined whether implemented code conforms to specifications that remain written in natural language without using dynamic testing. This limitation is particularly important in industrial safety and cybersecurity settings, where many safety- and security-relevant behaviors are still defined primarily in prose and where implementation correctness depends on the intended functional meaning of the requirement, not merely on local coding quality.

The limitation of conventional tooling is straightforward. Static analyzers such as Coverity and SonarQube are effective at identifying coding defects, unsafe patterns, and known vulnerability classes, but they do not establish whether implemented behavior matches the functional specification. Consider a simple example: if a program is intended to compute multiplication but the implementation performs addition, the code may contain no memory-safety problem, no suspicious API usage, and no recognizable vulnerability pattern. The defect is nonetheless real, and detecting it requires comparing implementation behavior with the specification. The same limitation appears in cybersecurity business logic: a requirement may state that a system-generated password shall not exhibit an identifiable pattern, while the implementation still imposes a regular structure such as a letter in the first position and a digit in the second. Conventional code-based scanners do not detect this kind of requirement-implementation inconsistency. In theory, dynamic testing could detect such defects; in practice, however, it is difficult due to the limited numbers of test cases used in testing. By contrast, the objective of the present work is the development of a requirement-aware, semantics-aware static analysis methodology that inspects source code against requirements specifications without requiring program execution.

This article presents an industrial case study and workflow for the above specification-dependent verification problem in intelligent-vehicle cybersecurity software. Our first AI agent, ruleMiner, analyzes requirements sentence by sentence and retains only logic that is both normative and falsifiable, producing a verifiable rule set while also exposing ambiguity, self-contradiction, unclear wording, and other non-verifiable content through a \texttt{requirements\_specs\_issues} data structure. Our second AI agent, codeAuditor, then evaluates source code against the extracted rules by reasoning over triggers, constraints, prohibited states, and cross-file evidence rather than relying on shallow keyword matching.

This two-stage design is a central methodological choice rather than an implementation convenience. Instead of asking a single LLM to judge directly whether code satisfies a long natural-language specification, the workflow introduces an explicit intermediate representation that makes the verification target narrower, more inspectable, and easier to audit. This separation helps control common LLM failure modes, including hallucination, output variability, limited explainability, and context loss, while preserving the industrial objective of connecting requirement clarification with implementation verification in a single assurance pipeline.

\section{Methodology}

The workflow is organized as a two-stage verification pipeline. In the first stage, natural-language requirements are transformed into a sanitized, verifiable rule set while explicitly preserving requirement-quality defects. To improve result completeness, the AI agent ruleMiner leverages the following metamorphic relation:

\begin{quote}
MR1: When the underlying LLM temperature is set to its minimum value, an ideal rule miner should generate consistent outputs for the same input across multiple executions.
\end{quote}

However, our observations show that even with the temperature fixed at 0, ruleMiner may still produce different outputs across repeated executions. Based on this observation, we adopt a pooling strategy that reconciles and, when necessary, merges the outputs from multiple runs to obtain a more complete set of mined rules. In our experiments, ruleMiner was executed between one and three times. Readers are referred to Chen et al.~\cite{chen2018} for a formal definition of metamorphic relations.

The second stage checks the resulting rules against implementation evidence in the code base.

The first and second stages are carried out by the LLM-based AI agents ruleMiner and codeAuditor, respectively, as summarized in Table~\ref{tab:workflow}.

\begin{table}[htbp]
\caption{Two-stage verification workflow}
\label{tab:workflow}
\centering
\footnotesize
\setlength{\tabcolsep}{4pt}
\resizebox{\columnwidth}{!}{%
\begin{tabular}{|l|l|l|}
\hline
\textbf{Input} & \textbf{Agent} & \textbf{Output} \\
\hline
Requirements document & ruleMiner & \shortstack[l]{Verifiable rule set +\\ \texttt{requirements\_specs\_issues}} \\
\hline
 \shortstack[l]{Verifiable rule set +\\ source code} & codeAuditor & \shortstack[l]{Consistency-checking\\ report} \\
\hline
\end{tabular}
}
\end{table}

On the requirements side, ruleMiner is instructed to preserve only enforceable logic, typically explicit ``shall'' statements and security-relevant ``should'' statements when the surrounding context makes them effectively mandatory. This focus on extracting operational obligations from normative text is consistent with prior work on analyzing regulatory and policy rules for requirements engineering~\cite{breaux2008}. Explanatory, permissive, illustrative, and subjective material is excluded from the rule base. Terms such as ``random,'' ``strong,'' ``lowest,'' or ``instantly'' are not converted into rules unless they can be grounded in measurable verification conditions. Likewise, when a requirement contains internal conflict, undefined operational states, or incomplete acceptance criteria, the problematic content is recorded in the data structure \texttt{requirements\_specs\_issues} rather than translated into a misleading rule. This separation is important because it prevents defective requirement text from silently contaminating downstream implementation checks.

The code-side stage builds on that cleaned rule set. codeAuditor evaluates each rule primarily through its semantic content, deriving concrete verification points such as allowed and prohibited values, minimum lengths, threshold counts, uniqueness conditions, and operational triggers. It then checks these points against source code, configuration, constants, validation logic, and cross-file call relationships. Representative code-centric security analyses, including code property graph approaches, are powerful for structural vulnerability discovery, but they are not designed to determine whether implementation semantics satisfy natural-language business rules~\cite{yamaguchi2014}. A rule is not treated as satisfied simply because a suggestive identifier, comment, or isolated check appears in the code. Conversely, a prohibition is not treated as violated merely because related functionality exists somewhere in the code base; the analysis instead seeks evidence of an actual enabling path, reachable behavior, or unsafe fallback. In this sense, the workflow is business-logic-driven rather than pattern-based.

\section{Industrial Case Study and Observations}

The source materials describe a public-safety industrial case involving an in-vehicle WiFi security subsystem. A representative example on the requirements side concerned password-policy text that required uppercase letters, lowercase letters, digits, and special characters, while also mistakenly providing an example pattern that did not actually satisfy those conditions. Instead of converting this statement into an apparently precise verification rule, ruleMiner recorded it in the \texttt{requirements\_specs\_issues} data structure as contradictory content. This distinction is important in practice: the pipeline does not conceal specification defects by over-normalizing them into rules that appear machine-checkable but are not semantically reliable. Notably, this requirement defect remained undetected for more than a year.

A second sanitized example illustrates how requirement-side uncertainty interacts with code-side evidence. One requirement stated that a WiFi hotspot should be disabled automatically when the system was ``not in use,'' but the triggering state was not defined precisely enough to support unambiguous verification from the specification alone. The requirement-side result was therefore marked as low-confidence rather than treated as a fully stable rule. Even so, the subsequent code audit reported a high-confidence mismatch: the implemented trigger depended on some local inactivity conditions instead of the whole system being ``not in use,'' while no evidence was found for integration with the broader operational state that the requirement appeared to imply. More generally, the analysis examines implemented trigger logic and reachable behavior rather than merely matching requirement phrases or suggestive identifiers. This reveals a requirement defect rather than coding bugs.

The quantitative results currently available should be interpreted as preliminary rather than as benchmark claims. On the requirement side, the broader analysis reported 75 requirement issues out of 222 analyzed items. For the reviewed WiFi security requirement subset, manual inspection found no incorrect extracted rules. Recall was not evaluated because exhaustive completeness checking at that scale would have required substantial additional manual effort.

On the code side (note that the code base was mature as it passed previous tests), analysis of the WiFi code base produced 8 Fail/Unknown findings. Among the manually inspected findings, no false positives were observed, and one high-priority coding issue was detected and later fixed by the development team. The few disputable cases were not attributed to failures of the verification technology itself; rather, they arose from incomplete requirement-to-implementation mapping and missing cross-department traceability information, which limited definitive interpretation of some findings. This distinction matters because it helps separate gaps in organizational traceability from limitations of the underlying verification method.

\section{Discussion and Lessons Learned}

Three lessons emerge from this study. First, requirements analysis should explicitly model defects in the requirements text itself. The \texttt{requirements\_specs\_issues} data structure is essential because it separates enforceable logic from statements that are contradictory, ambiguous, or insufficiently specified; without this separation, automated checking can create false confidence.

Second, business-logic-oriented code auditing is often necessary for cybersecurity requirements whose meaning depends on operational context. In practice, important defects arise not only from missing keywords or local syntactic violations, but also from mismatched triggers, partial enforcement, or incorrect execution paths. The most useful analyses therefore reason about intended behavior rather than textual resemblance alone. This is particularly relevant for defects that are difficult to expose through testing, especially when suitable test oracles are hard to define or when meaningful evaluation would require many generated samples~\cite{barr2015}. A semantics-aware static-analysis methodology can therefore complement dynamic testing by surfacing requirement-implementation inconsistencies earlier in the development lifecycle, without requiring compilation, execution, or a target runtime environment. The objective is not to replace testing, but to introduce a shift-left verification capability that can reduce testing effort for certain classes of defects while also identifying issues that conventional code-based scanners are not designed to detect. In this case study, the proposed static-analysis pipeline was able to verify more than 50\% of the requirements that previously could only be evaluated through software testing. In addition, the approach may help mitigate the test oracle problem~\cite{barr2015}, because the LLM-based analysis toolchain reasons about source-code semantics and implementation logic rather than relying solely on expected program outputs.

Third, requirements analysis and implementation analysis are most effective when treated as a coupled process. Requirements are increasingly consumed not only by human engineers but also by AI-assisted workflows, which can be less robust to ambiguity, contradiction, and missing traceability. An important industrial lesson, therefore, is that requirements should be clear, unambiguous, and non-contradictory, and that mappings between requirements and code bases---especially merge-request and change-history traceability---should be complete. These concerns align with longstanding observations in natural-language-processing-based requirements engineering, particularly around ambiguity management and traceability~\cite{dalpiaz2018,zhao2021}. This dependency on high-quality traceability is highly visible in industrial settings but is often underemphasized in academic discussions.

\subsection{Future Work}

Current quality control on the requirements side relies on the rule-sanitization criteria described above and on manual inspection of the WiFi security requirement set examined in this report. ruleMiner preserves only enforceable logic and records contradictory, ambiguous, or otherwise non-verifiable requirement content rather than translating such content into rules. The LLM used by our agents was Claude Opus 4.6.

As future work, metamorphic relations and metamorphic testing could be further applied to rule mining. One straightforward approach is to submit semantically equivalent rewrites or paraphrases of the same requirement, then compare and, where appropriate, merge the resulting rule sets. Additional metamorphic relations could involve the addition or deletion of relevant statements to verify whether the mined rules change accordingly, thereby providing evidence for the correctness and effectiveness of ruleMiner. Simlar strategies may be applied to codeAuditor, too.

\section{Conclusion}

This paper reports an industrial LLM-assisted workflow for detecting business-logic errors in cybersecurity requirements and implementations. The central idea is to couple verifiable rule extraction with the explicit recording of \texttt{requirements\_specs\_issues}, followed by code auditing that reasons about business logic rather than relying on shallow pattern matching. Metamorphic relations have been used to enhance the output quality of the toolchain. The reported case materials show that the workflow can precisely expose a large number of contradictory or non-verifiable requirements, support rule extraction, and identify implementation-level defects, including a high-priority bug in a mature WiFi code base that was subsequently fixed. Recall and completeness of rule mining were not evaluated in this study.

Although the current evidence remains preliminary and limited in scope, the reported experience suggests that integrating requirements clarification, traceability awareness, and LLM-based requirements-aware static analysis within a unified workflow represents a promising practical direction for industrial assurance. More specifically, the workflow should be understood as complementing, rather than replacing, dynamic testing and conventional static-analysis tools. It supports shift-left verification by requiring no compilation, execution, or runtime environment, and it may reduce testing effort for classes of defects whose manifestation depends on specification semantics and the availability of suitable test oracles, making them difficult to expose through conventional test cases. Here, ``specification semantics'' refers to the intended constraints and business logic expressed by the requirements, rather than their surface wording alone.

\balance

\end{document}